\documentclass[useAMS]{mn2e}

\usepackage{graphicx, natbib}

\newcommand{\aap}{A\&A}

\newcommand{\aj}{AJ}
\newcommand{\apj}{ApJ}
\newcommand{\apjl}{ApJL}
\newcommand{\apjs}{ApJS}

\newcommand{\araa}{ARA\&A}
\newcommand{\mnras}{MNRAS}
\newcommand{\nat}{Nat}

\newcommand{\iaucirc}{IAU Circ.}

\newcommand{\groj}{GRO~J\,1744-28}

\newcommand{\ks}{K_{\rm S}}
\newcommand{\brg}{Brackett-$\gamma$}

\newcommand{\ltsim}{\raisebox{-0.6ex}{$\,\stackrel
        {\raisebox{-.2ex}{$\textstyle <$}}{\sim}\,$}}
\newcommand{\gtsim}{\raisebox{-0.6ex}{$\,\stackrel
        {\raisebox{-.2ex}{$\textstyle >$}}{\sim}\,$}}
\newcommand{\Rsol}{{\rm R_{\odot}}}
\newcommand{\Msol}{{\rm M_{\odot}}}

\newcommand{\chandra}{{\it Chandra}}
\newcommand{\xmm}{{\it XMM-Newton}}
\newcommand{\ukidss}{{\it UKIDSS}}
\newcommand{\vista}{{\it VISTA}}
\newcommand{\tmass}{{\it 2MASS}}

\newcommand{\jh}{J\!-\!H}
\newcommand{\jk}{J\!-\!K}
\newcommand{\hk}{H\!-\!K_{\rm S}}

\title[GRO J1744-28, search for the counterpart]{GRO J1744-28, search for the counterpart: infrared photometry and spectroscopy}

\author[A. J. Gosling et al.]{A.J. Gosling,$^{1}$\thanks{e-mail:
ajg@astro.ox.ac.uk} R.M. Bandyopadhyay,$^{2}$ J.C.A. Miller-Jones$^{3}$ and S.A. Farrell,$^{4,5}$ \\
$^{1}$Department of Astrophysics, University of Oxford, Keble Road,
Oxford, OX1 3RH, UK \\
$^{2}$Department of Astronomy, University of Florida, Gainesville, FL 32611, USA\\
$^{3}$Astronomical Institute `Anton Pannekoek', University of
Amsterdam, Kruislaan 403, 1098 SJ Amsterdam, The Netherlands\\
$^{4}$School of PEMS, UNSW@ADFA, Northcott Drive, Canberra, ACT 2600, Australia\\
$^{5}$Centre d'Etude Spatiale des Rayonnements, 9 Avenue du Colonel Roche, 31028 Toulouse Cedex 4, France CNRS/UPS}

\begin{document}

\date{\today}

\pagerange{\pageref{firstpage}--\pageref{lastpage}} \pubyear{2007}

\maketitle


\label{firstpage}

\begin{abstract}
\noindent

Using VLT/ISAAC, we have detected 2 candidate counterparts to the
bursting pulsar \groj, one bright and one faint, both within the X-ray
error circles found using {\it XMM-Newton} and {\it Chandra}. In
determining the spectral types of the counterparts we applied 3
different extinction corrections; one for an all-sky value, one for a
Galactic Bulge value and one for a local value. We find that the local
value, with an extinction law of $\alpha = 3.23 \pm 0.01$ is the only
correction that results in colours and magnitudes for both the bright
and faint counterparts that are consistent with a small range of
spectral types, and in the case of the bright counterpart are also
consistent with the spectroscopic identification.  Photometry of the
fainter candidate then indicates it is a K7/M0~V star at a distance of
$3.75\pm1\,{\rm kpc}$. Such a star would require a very low
inclination angle ($i<9^{\circ}$) to satisfy the mass-function
constraints; however this source cannot be excluded as the counterpart
without follow-up spectroscopy to detect emission signatures of
accretion.  Photometry and spectroscopy of the bright candidate
indicate that it is most likely a G/K~III star. The spectrum does not
show \brg\ emission, a known indicator of accretion. The bright star's
magnitudes are in agreement with the constraints placed on the
probable counterpart by the calculations of \citet{rapp97} for an
evolved star that has had its envelope stripped.  The mass-function
indicates the most likely counterpart has $M < 0.3\,\Msol$ for an
inclination of $i \geq 15\degr$; a stripped giant, or a main
sequence M3+~V star would be consistent with this mass-function
constraint.  In both cases mass-transfer, if present, will be by
wind-accretion as the counterpart will not fill its Roche lobe given
the observed orbital period. In this case, the derived magnetic field
strength of $2.4 \times 10^{11}$ G is sufficient to inhibit accretion
of captured material by the propeller effect, hence the quiescent
state of the system.

\end{abstract}



\section{Introduction}

The Bursting Pulsar \groj, was discovered with the Burst and
Transient Source Experiment (BATSE) on the {\it Compton Gamma-Ray
Observatory} ({\it CGRO}) on $2^{\rm nd}$ December 1995
\citep{fish95, kouv96} during a period of outburst. \groj\ was
only the second system (after the Rapid Burster MXB 1730-335) to
exhibit Type-II X-ray bursts. \groj\ displays properties of both a
pulsar and a Type-II burster, making it a unique source for
studying the properties of accretion onto neutron stars (NS) and
the interaction of magnetic fields with accretion flows.

\citet{fing96} detected coherent X-ray pulsations with a period of
467\,ms. The pulsation rate increased during their observations at a
rate of $1.2 \times 10^{-11}$\,Hz\,s$^{-1}$.  They were able to
determine the orbital period of the system to be $P_{\rm orb} =
11.8337 \pm 0.0013$\,days by fitting pulse phases.  Determining
$P_{\rm orb}$ through phase shifting, which is a result of Doppler
shifts of the pulse period due to orbital motion indicates that it is
unlikely that the system is face-on.  These measurements indicate that
the system consists of a magnetised NS with a magnetic field $>
10^{11}$ G \citep{gile96, lewi96}, and that the accretion is spinning
up the NS. \citet{cui97} also measured the pulsations of \groj\ and
was able to derive a magnetic field strength of $2.4 \times
10^{11}\,{\rm G}$, assuming that the source was entering the propeller
regime at the end of its phase of outburst.


The system has undergone two periods of outburst; one lasting from
$2^{\rm nd}$ December 1995 until $26^{\rm th}$ April 1996, the second
lasting from $1^{\rm st}$ December 1996 until $7^{\rm th}$ April 1997
\citep{wood99}. During these periods of outburst, hard X-ray bursts
were initially observed with durations typically between 8 and
$30$\,s, and a frequency of $\sim 20$ per hour, falling to $\sim 1$
per hour after 1 day \citep{wood99}. It is believed that such
outbursts are caused by instabilities in the accretion disc, leading
to short intervals of increased accretion onto the surface of the NS
\citep{lewi96}.


\begin{center}
\begin{figure*}
\scalebox{0.25}{\includegraphics{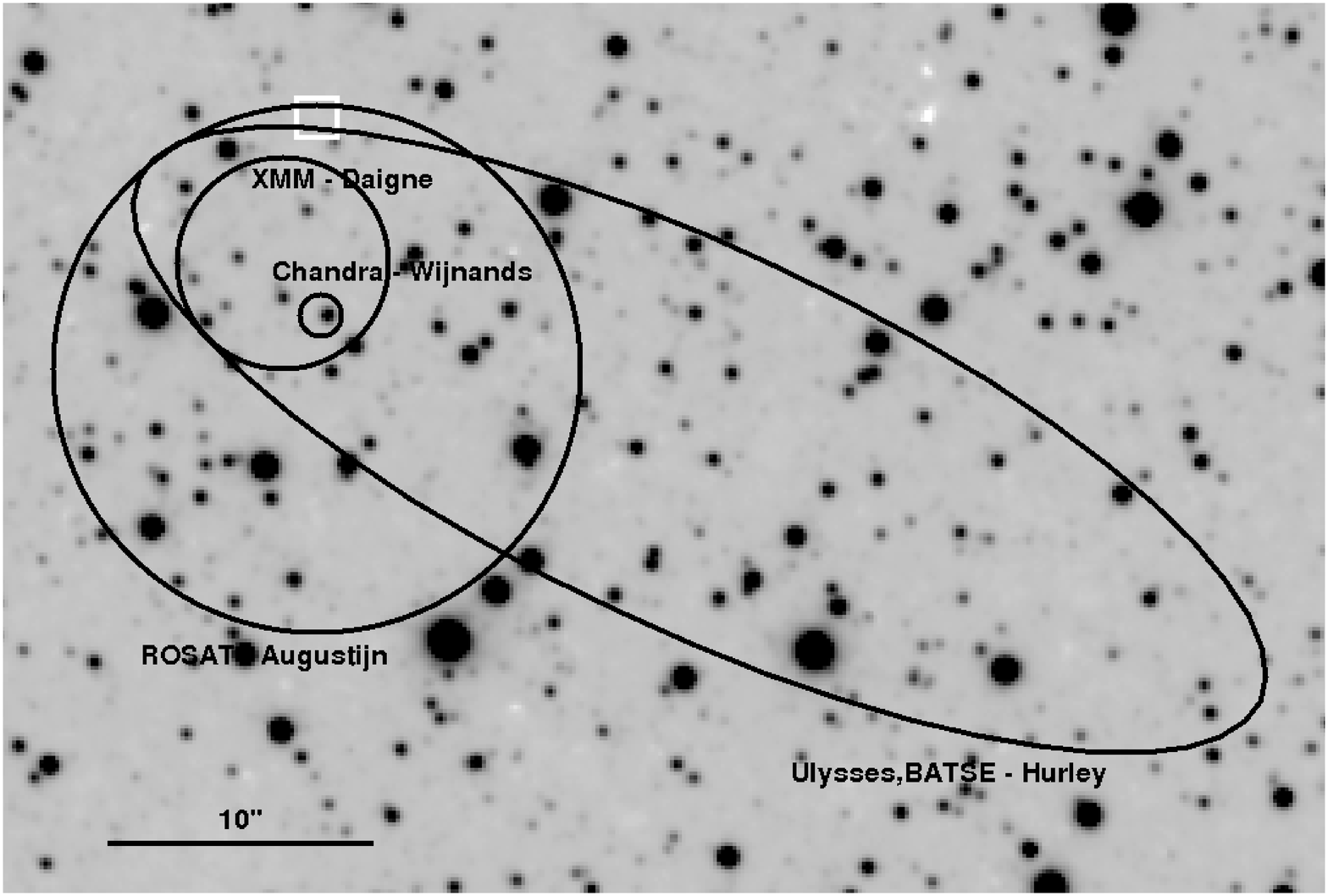}}
\scalebox{0.4}{\includegraphics{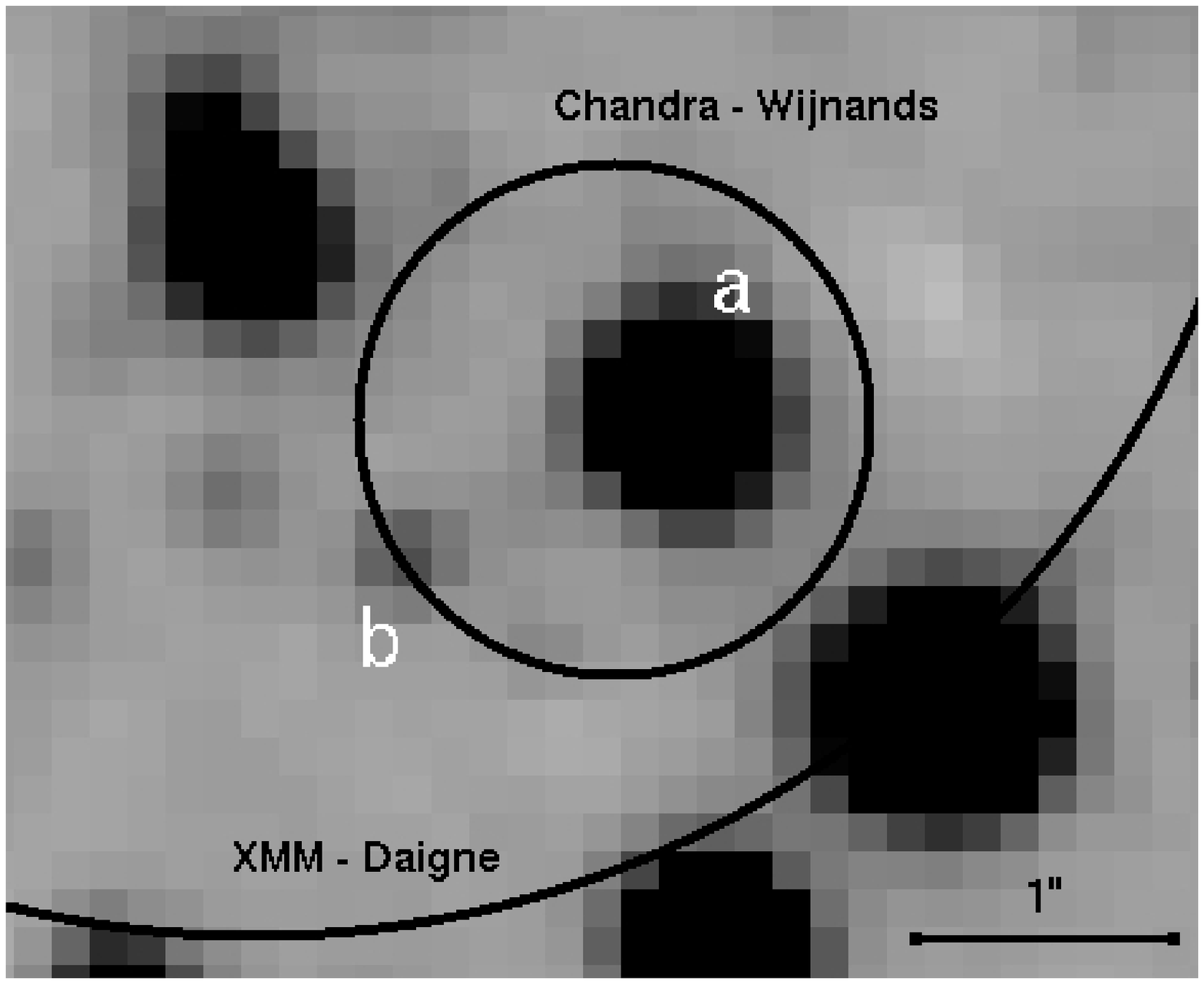}}
\caption{{\it left}: Position error circles for \groj\ as determined
by different X-ray instruments. These are overlaid on our $\ks$-band
image taken with VLT. The white square near the top of the image is
the position error of the previously claimed counterpart
\citep[$R.A. = 17^{\rm h}44^{\rm m}33\fs1 \pm 0\fs1$, $Dec. =
-28^{\circ}44\arcmin19\farcs5 \pm 1\farcs5$,][]{augu97,cole97}. {\it
right}: Close up of the \chandra\ and \xmm\ error circles over the
$\ks$-band VLT image.  The star labelled {\it a} is the bright
candidate counterpart for which we obtained a spectrum. The source
labelled {\it b} is the faint candidate counterpart for which we were
unable to obtain a spectrum in our VLT data.}
\label{finder}
\end{figure*}
\end{center}


In 2001, \groj\ was in a quiescent state with a very low level of
X-ray emission. Its $0.5-10\,{\rm keV}$ luminosity was measured to be
$2-4 \times 10^{33}$\,erg\,s$^{-1}$ with \chandra\ \citep[assuming the
source is near the Galactic Centre (GC) at a distance of $8\,{\rm
kpc}$;][]{wijn02}. More recently, \citet{muno07} re-observed
\groj. Its $2-8\,{\rm keV}$ luminosity was measured to be $6 \times
10^{33}$\,erg\,s$^{-1}$, indicating increased activity from the X-ray
source. These measurements are within the observed range of quiescent
X-ray transients which exhibit luminosities of $10^{30} -
10^{34}$\,erg\,s$^{-1}$ in these bands.

In this paper we describe the characteristics of two astrometrically
selected candidate counterparts to the X-ray source and find that both
are unlikely to be the true counterpart. From this we derive
constraints as to the structure and composition of the binary system
\groj.

\section{Position}

The accuracy of the measured position of \groj\ has improved
drastically over time as the astrometric precision of X-ray telescopes
has improved. Initially, the position was only available with an error
radius of $\sim 6^{\circ}$ at $1\,\sigma$ \citep{fish95}. However, this
was quickly refined and as new instruments and techniques were
developed \groj\ has been revisited many times. The positional error
has been a critical issue for those searching for the
stellar counterpart to \groj, as the GC has a very high stellar
density, with an average stellar separation $\leq 1.94$\arcsec\ in the
$\ks$-band at a magnitude limit of $\ks=20$ \citep{gosl06}. The
current best measurement of the X-ray position is that of \citet{wijn02},
obtained using \chandra.

\cite{cole97} and \citet{augu97} announced the discovery of an
optical/near-infrared counterpart candidate based on the {\it ROSAT}
position for \groj. The positions reported by both groups were
consistent within their errors. This counterpart was reported to show
variability as it was present in only some of the observations for
each telescope used. This candidate star was at the edge of the {\it ROSAT}
error circle. More recently {\it XMM-Newton} \citep{daig02} and
\chandra\ \citep{wijn02} observations of \groj\ have shown that this
variable star cannot be the counterpart to \groj\ as these new, more
accurate observations of the X-ray position are not coincident with
the position of the previously proposed counterpart (see
Fig. \ref{finder}).


\section{Data and Reduction}

As part of a program of follow up study of the X-ray sources
discovered by \citet{wang02}, we have obtained IR imaging and
spectroscopy of likely counterparts to the X-ray sources. The imaging
is intended to select candidate counterparts astrometrically, and
determine the colours of the general population of counterparts in
comparison to the field population. To conclusively identify the
counterparts to the X-ray sources, we are also undertaking a program
of spectroscopic observations to identify accretion signatures in
their spectra \citep{band97, band99}.

The observations of \groj\ upon which this paper is based were were
carried out in 2003 and 2005 during which period \groj\ was assumed to
be in quiescence based on X-ray luminosities measured by \chandra\ in
2001 and 2007 \citep{wijn02,muno07} and also the fact that no
outbursts were detected by any of the X-ray all-sky monitors.


\begin{center}
\begin{figure*}
\includegraphics[width=120mm]{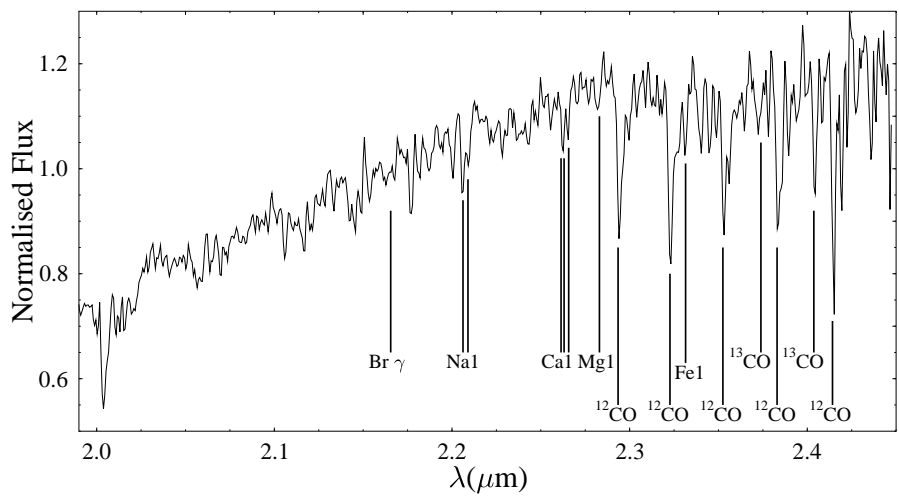}
\caption{Combined spectrum of the bright candidate counterpart to
\groj\ from the two nights of observation. The prominent absorption
lines have been indicated. The strong CO lines indicate that this
source is a late type giant. The lack of \brg\ (line position
indicated although not present) indicates that the star is not the
X-ray source counterpart, or that the accretion signature is too weak
to be distinguished from the noise.}
\label{speccomb}
\end{figure*}
\end{center}


\subsection{Imaging}
\label{sec:imaging}

In June and July of 2003, we observed 26 fields within the GC \chandra\ mosaic
of \citet{wang02} using ISAAC, a $1024 \times 1024$ pixel Hawaii
Rockwell detector on the ESO VLT. This provides a $2.5\arcmin \times
2.5\arcmin$ field of view on the sky with $0\farcs1484$ resolution per
pixel. One of these locations was chosen to cover the region
containing \groj. This region was observed on the 26$^{\rm th}$ July
2003. We obtained 6 minutes of exposure per pointing in each of the
three near-IR bands, $J$, $H$ and $\ks$, on nights with seeing $\ltsim
0.6$\arcsec. The average magnitude limits of the images are
$J=23~(S/N=5)$, $H=21~(S/N=10)$ and $\ks=20~(S/N=10)$. The initial reduction
(flat-fielding, removal of bad pixels, and sky subtraction) was
performed with the ESO/ISAAC pipeline reduction software. These image
products were then astrometrically locked to the Two Micron All-Sky
Survey (2MASS) resulting in a $0\farcs1$ positional
uncertainty. Photometric calibration was first carried out using the
VLT zero-points and observed photometric standards. The data were then
compared to the 2MASS catalogue and a photometric offset was applied
to transform the VLT system to the photometric system of 2MASS. Source
positions and magnitudes were measured using SExtractor \citep[version
2.3.2; see][for further details]{band05}. We then performed
astrometric matching of the IR source positions to the X-ray source
positions to identify possible counterparts to the X-ray sources.

We detected two sources in all three bands within the error circle of
the \emph{Chandra} position for \groj.  The position of the brighter
of the two sources is $R.A. = 17^{\rm h}44^{\rm m}33\fs07$, $Dec. =
-28^{\circ}44\arcmin26\farcs89$ with a $0\farcs1$ error. It has
reddened magnitudes of $J = 18.98 \pm0.02$, $H = 15.68 \pm 0.01$ and
$\ks = 14.44 \pm 0.01$.  The position of the fainter of the sources is
$R.A. = 17^{\rm h}44^{\rm m}33\fs16$, $Dec. =
-28^{\circ}44\arcmin27\farcs41$ with a $0\farcs1$ error. It has
reddened magnitudes of $J = 21.5 \pm 0.1$, $H = 19.0 \pm 0.1$ and $\ks
= 18.3 \pm 0.1$.

\subsection{Spectroscopy}

We obtained $\ks$ spectra for the candidate IR counterparts to 27 of
the X-ray sources, identified in the imaging program described in
Section~\ref{sec:imaging}. Spectra were obtained using the long slit
mode on VLT/ISAAC with a 1\arcsec\ slit-width, R = 450, in service
mode in period 75 (April - September 2005). O and B type standard
stars were observed for each spectrum. The spectra for the brighter of
the two candidate counterparts to \groj\ were obtained on the $2^{\rm
nd}$ July and $4^{\rm th}$ September, each with 120 s integrations,
giving a total integration time on source of 240\,s.

The spectra were reduced using the IRAF routine ``apsum'' to extract
the 1 d spectra. Atmospheric lines were removed by dividing by the
standard star spectra using the IRAF routine ``telluric''. The spectra
were observed on non-photometric nights to increase the chances of the
observations being performed so no flux standards were observed,
therefore they cannot be flux calibrated. The spectra have been
normalised to 1 by dividing by the mean of the flux values. Identified
lines and equivalent widths measured in the combined spectrum of the
brighter candidate star are listed in Table \ref{widths}, and the
spectrum is shown in Fig.~\ref{speccomb}.

\begin{table}
\begin{center}
\begin{tabular}{lll}
\hline
Line & $\lambda$ & Width\\
 & ($\mu m \pm 0.002$) & (\AA) \\
\hline
Na I (doublet)  & $2.207$ & $5.98 \pm 0.1$\\
Ca I (triplet)  & $2.264$ & $4.25 \pm 0.1$\\
Mg I            & $2.282$ & $1.73 \pm 0.1$\\
$^{12}$CO (2-0) & $2.295$ & $7.43 \pm 0.8$\\
$^{12}$CO (3-1) & $2.323$ & $9.71 \pm 0.6$\\
Fe I            & $2.331$ & $2.07 \pm 0.1$\\
$^{12}$CO (4-2) & $2.353$ & $6.77 \pm 0.3$\\
$^{12}$CO (3-1) & $2.372$ & $2.43 \pm 0.1$\\
$^{12}$CO (5-3) & $2.383$ & $9.43 \pm 1.2$\\
$^{13}$CO (4-2) & $2.404$ & $5.22 \pm 0.2$\\
$^{12}$CO (6-4) & $2.415$ & $10.38 \pm 1.1$\\

\hline
\end{tabular}
\caption{Identified lines and measured equivalent widths from the
combined spectrum of the bright source (Fig.~\ref{speccomb}).}
\label{widths}
\end{center}
\end{table}


\section{Extinction}
\label{ext}

The original, and generally applied, extinction law was calculated in
the 1980's using relatively primitive (by today's standards) infrared
detectors, using a small sample of individually selected stars whose
spectral types were well known \citep{riek85, card89, math90, catc90}.
Based on this extremely small sample, they were able to express the
near-infrared extinction law as ``universal'', a fact that has been
adopted in almost all studies since. However, in the last decade, as
infrared detector technology has advanced, we have been able to obtain
much higher-resolution, deeper surveys of the sky in the infrared band (\tmass,
{\it DENIS}, \ukidss, \vista). Based on these observations, it has
become evident that the extinction law is not the ``universal'' value
that it was previously thought to be, but is in fact highly variable
from point-to-point \citep[][and others]{mess05, nish06, fitz07,
froe07}. As we showed in \citet{gosl06}, there is a high degree of
spatial variation in the extincting material towards the Bulge, and we
have also discovered that there is variation in the extinction law
towards the Bulge (Gosling et al. {\it in prep.}). \citet{fitz07}
found that the Galactic extinction curve varies greatly from
sight-line to sight-line and that any extinction curve will reflect
biases in the local population. They also concluded that it is difficult
to meaningfully characterise average extinction properties due to the
sensitivity of the extinction to local conditions.

It is for this reason that in this paper we present extinction
corrections, and the possible candidate counterpart properties that
each imply, for the canonical value, a value determined by
\citet{nish06} that deals specifically with the Galactic Bulge, and a
more specific calculation based on the work of Gosling et al.\ ({\it
in prep.}).  We have presented all three so that the effect of the
extinction correction on the determination of the spectral types of
the candidate counterparts can be compared and to highlight the
importance of making the correct extinction correction in the analysis
of any source within this region.  The derived extinction values are
summarised in Table~\ref{extinctval}.

We show that only a specific, local extinction correction that takes
into account the possibility for the extinction law to vary from the
perceived universal value can produce consistent colours and
magnitudes for both the bright and faint candidate counterparts, and
that the results of this are entirely consistent with the possible
spectral types derived from analysis of the spectrum of the brighter
of the two candidate counterparts.


\begin{center}
\begin{figure}
\includegraphics[width=\columnwidth]{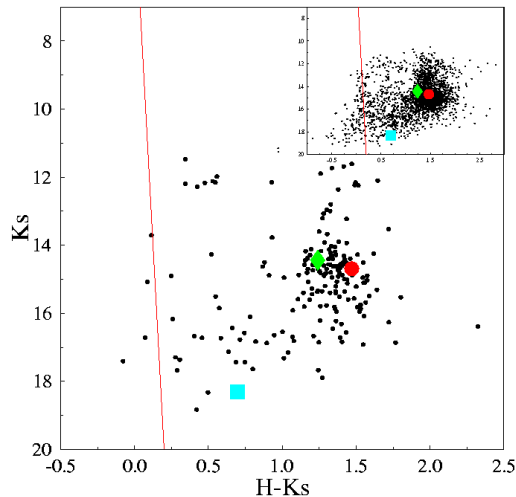}
\caption{The main diagram is the $\hk$ vs $\ks$ colour-magnitude
diagram for the stars extracted within 20\arcsec\ of \groj; the
smaller diagram inset is for the whole field, shown for comparison.
The red line is the loci of the foreground stars that was fit after
finding maxima in the colour distribution for different magnitudes as
described in the text. The green diamond indicates the reddened values
for the bright candidate counterpart from this work; the red circle corresponds to
the magnitudes of the same star as reported by \citet{wang07} for
comparison. The blue square is the position of the faint candidate
counterpart.}
\label{colmag}
\end{figure}
\end{center}



\begin{center}
\begin{figure}
\includegraphics[width=\columnwidth]{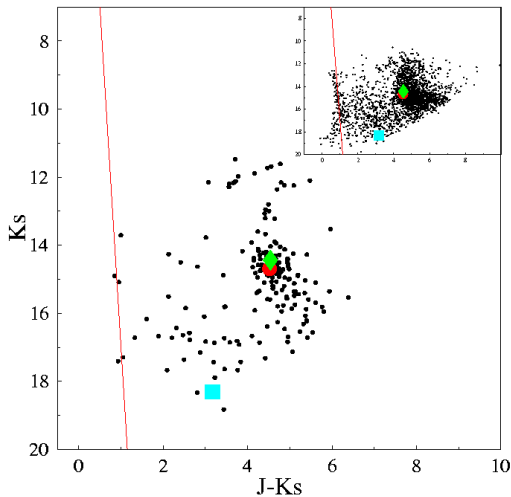}
\caption{$\jk$ colour-magnitude diagram for the stars surrounding
\groj\ and (inset) for the whole field. The markings are the same as those
in Figure \ref{colmag}. Both the bright and faint sources are
consistent with the general population in the field, and the bright
source observed by \citet{wang07} has colours consistent with those
observed in this work.}
\label{colmagjk}
\end{figure}
\end{center}


\subsection{Canonical Value - all sky}
\label{extall}

\citet{dota96a} measured the column density to \groj\ to be $5.3\pm0.1
\times 10^{22}\,{\rm cm^{-2}}$ for the persistent emission, and
$6.1\pm0.2 \times 10^{22}\,{\rm cm^{-2}}$ for the
outbursts. \citet{muno07} measured the column density to be $6.5\pm0.1
\times 10^{22}\,{\rm cm^{-2}}$.  Using the formula from \citet{bohl78}
we can convert the value for the persistent emission (since the source
is not in a bursting period) to $A_V =
28.3\,{\rm mag}$. This is close to the canonical value for extinction
towards the Galactic Bulge of $A_V = 25-30\,{\rm mag}$. Using the
conversion factors from \citet{riek85}, this $A_V$ becomes $A_J =
7.98$, $A_H = 4.95$ and $A_{\ks} = 3.20$.


\subsection{Nishiyama Value - Galactic Bulge}
\label{extgb}

We extracted all the stars from within 20\arcsec\ of the positions of
the counterparts for which we had three-colour information, a total of
192 stars, in order to calculate the average colour excess for the
counterparts. We obtained mean colours and magnitudes for the stars
surrounding the candidate counterparts to \groj\ of $J = 19.34$, $H =
16.10$ and $\ks = 14.79$ and $\jh = 3.22$, $\jk = 4.58$ and $\hk =
1.35$. We obtained lines representing the foreground branch in each of
the colour-magnitude diagrams (the population to the left of each
diagram) by fitting the maximum in the colour distribution around the
left hand population at different magnitudes for all the stars in the
field (see Figures \ref{colmag} and \ref{colmagjk}). Using this, we
obtain the colour excesses to the stars surrounding the candidate
counterparts to be $E\!(\!\jh\!) = 2.48$, $E\!(\!\jk\!) = 3.69$ and
$E\!(\!\hk\!) = 1.09$. As the stellar population has the least spread
in colour in the $\ks\,vs\,\hk$ colour-magnitude plane, we will use
the excess from this for calculating the absolute extinctions as it
introduces the lowest error due to intrinsic scatter in the
colour. The error in the determination of the colour excess in $\hk$
is $0.055$.

\citet{nish06} analysed the stars of the Galactic Bulge and observed a
variation in the relationship of extinction to wavelength across the
quadrants of the North-East, North-West, South-East and South-West
areas of the Bulge (with respect to Sgr A*). Extracting the red-clump
stars from their data, and plotting their colours they were able to
measure the extinction law values for these quadrants. \groj\ lies in
the quadrant they label N+, which has a measured slope for the
extinction law of $\alpha = 1.96 \pm 0.01$
($A_{\lambda}\propto\lambda^{-\alpha}$), which equates to extinction
relations of $A_J : A_H : A_K = 1 : 0.590 \pm 0.009 : 0.342 \pm
0.004$. Converting the colour-excess to absolute extinction with the
formula $$ A_{\lambda_{2}} = \frac{ \left< E( \lambda_{1} \! - \!
\lambda_{2} ) \right> }{ \left( \frac{ \lambda_{2} }{ \lambda_{1} }
\right) ^ {\alpha} - 1} $$ gives values of $A_J = 4.58 \pm 0.24$, $A_H
= 2.70 \pm 0.15$ and $A_{\ks} = 1.57 \pm 0.08$.


\subsection{Gosling Value - local}
\label{extloc}

We analysed the colour distribution of the stars in the field
containing \groj, using the same stars extracted for the Galactic
Bulge extinction calculation (see Section \ref{extgb}) and using a
method we have developed for correction of our own VLT data (Gosling
et al. {\it in prep.}), similar to that of \citet{froe06} we were able
to extract the slope of the extinction law for this specific
region. Using the ratio of the mean colours for the 192 stars within
20\arcsec of \groj, we used the formula $$ \frac{\left<
E(\lambda_{1}\!-\!\lambda_{2}) \right>}{\left<
E(\lambda_{2}\!-\!\lambda_{3}) \right>} =
\frac{\left(\frac{\lambda_{2}}{\lambda_{1}}\right)^{\alpha} - 1}{1 -
\left(\frac{\lambda_{2}}{\lambda_{3}}\right)^{\alpha}} $$ and its
permutations to obtain a value of $\alpha$ for $\left< E(\jk)
\right>/\left< E(\jh) \right>$, $\left< E(\jk) \right>/\left< E(\hk)
\right>$ and $\left< E(\jh) \right>/\left< E(\hk) \right>$.

Calculating the extinction law for all three colour-colour ratios gave
a value of $\alpha = 3.23 \pm 0.01$ in each case, which equates to
extinction relations of $A_J : A_H : A_K = 1 : 0.419 \pm 0.009 : 0.171
\pm 0.004$. Converting the colour-excess to extinction as in the
Galactic Bulge case gives $A_J = 4.60 \pm 0.27$, $A_H = 1.93 \pm 0.12$
and $A_{\ks} = 0.79 \pm 0.04$, all substantially lower than implied by
the other methods (with the exception of the $A_J$ value found for the
Galactic Bulge extinction scenario).

\begin{table}
\begin{center}
\begin{tabular}{lcccc}
\hline
               & $\alpha$        & $A_{J}$ & $A_{H}$ & $A_{\ks}$\\
\hline
all-sky        &                 & 7.98    & 4.95    & 3.20     \\
Galactic Bulge & $1.96 \pm 0.01$ & 4.58    & 2.70    & 1.57     \\
local          & $3.23 \pm 0.01$ & 4.60    & 1.93    & 0.79     \\
\hline
\end{tabular}
\caption{The values of absolute extinction in magnitudes for the three
possible types of extinction correction. Where appropriate the value
of $\alpha$, the extinction law slope is also given.}
\label{extinctval}
\end{center}
\end{table}

\section{Bright star}

\subsection{Star Properties}

Using the extinction values found in Section~\ref{ext} and summarised
in Table~\ref{extinctval}, we can determine for each case the possible
spectral types and luminosity classes of the bright candidate
counterpart.  The results, which we now go on to discuss in detail,
are summarised in Table~\ref{brightposs}.

\subsubsection{Corrected for all-sky extinction}

Correcting for the all sky extinction values, the intrinsic magnitudes
of the bright source become $J = 11.00 \pm 0.20$, $H = 10.73 \pm 0.15$
and $\ks= 11.24 \pm 0.10$ which gives colours of $\jh = 0.70 \pm
0.25$, $\jk = -0.24 \pm 0.22$ and $\hk = -0.51 \pm 0.18$. The $\hk$
colour is inconsistent with any spectral type, indicating that there
is either an error with the photometry, or with the extinction
correction (see Section \ref{sec:extinction} for discussion of this).

Within the errors, the $\jh$ colour is consistent with an O-F~V/I
star, and the $\jk$ colour is consistent with O-B4~V/I stars.

Correcting for a Galactic Centre distance of $7.5 \pm 0.45\,{\rm kpc}$
\citep{nish06b}, the magnitudes correspond to B0~V or K5/M0~III stars
\citep{alle00}.

A supergiant at the GC would appear 3--4 magnitudes brighter than
observed, or would have to be on the far side of the Bulge for its
apparent magnitude to be that observed. The colours are not consistent
with a giant star, so the candidate counterpart cannot be the
K5/M0~III star that is a possible match to the magnitudes. Only a B0~V
star is consistent with all the magnitudes and the $\jh$ and $\jk$
colours. The $\hk$ colour is not consistent with any spectral
type.


\subsubsection{Corrected for Galactic Bulge extinction}

Using the Galactic Bulge extinction values from Section \ref{extgb},
the intrinsic magnitudes of the bright source are $J = 14.40 \pm
0.24$, $H = 12.98 \pm 0.15$ and $\ks= 12.87 \pm 0.08$, giving the
bright star colours of $\jh = 1.42 \pm 0.28$, $\jk = 1.53 \pm 0.25$
and $\hk = 0.11 \pm 0.17$.

Of the colours, only the $\hk = 0.11 \pm 0.17$ is a realistic value
for a star, however, within the error bounds, all spectral types are
possible \citep{alle00}. Both the $\jh$ and $\jk$ colours are too
large for any spectral types, even at the limits of their errors. This
indicates that the $J$ magnitude is too faint (see Section
\ref{sec:extinction} for further on this).


Taking the $\hk = 0.11$ colour to be correct, the most probable
spectral types for this candidate include G8~I, K1~III and K4/5~V
\citep{alle00}. 

As the $J$-band magnitude seems to be too faint, we will only use the
$H$ and $\ks$ magnitudes of the star, which for a Galactic Centre
distance of $7.5 \pm 0.45\,{\rm kpc}$ \citep{nish06b} correspond to a
G6-K1~III or B2-4~V \citep{alle00}.

As with the all-sky correction, the supergiant type star consistent with the
colours would be 3--4 magnitudes brighter than observed if it is at a GC
distance, and would be on the far side of the Bulge. The K4/5~V star
would be too faint at a GC distance suggested by the star's position on
the colour-magnitude diagrams, for it to be this star it would have to
be at a distance of $\sim500$\,pc, very close by. The remaining option from the
$\hk$ colour is the K1~III which is also consistent with the $H$ and
$\ks$ magnitudes. We note again that the $\hk$ errors mean that any
spectral type is possible.

\subsubsection{Corrected for local extinction}
\label{brightloc}

Using the local extinction values from Section \ref{extloc},
the intrinsic magnitudes of the bright source are $J = 14.38 \pm
0.27$, $H = 13.76 \pm 0.12$ and $\ks= 13.65 \pm 0.04$, giving the
bright star colours of $\jh = 0.63 \pm 0.29$, $\jk = 0.73 \pm 0.27$
and $\hk = 0.10 \pm 0.13$. 

For this extinction correction, all three colours are consistent with
the same classes of stars within the errors, being G--M~V stars,
M2 or earlier giants and G/K~I stars. The most likely matches are
K4-6~V, K0-2~III or G3~I which are good fits for all $\jh$, $\jk$ and
$\hk$ colours \citep{alle00}. 

Correcting for a Galactic Centre distance of $7.5 \pm 0.45 \,{\rm kpc}$
\citep{nish06b}, the $J$, $H$ and $\ks$ magnitudes are consistent with
B5~V and G4~III stars \citep{alle00}.

As with the previous two cases, the supergiant solution to the colours
would be 3--4 magnitudes brighter than observed, or would have to be on
the far side of the Bulge. The K4-6~V star selected from the colours
would be too faint for the observed magnitudes unless the distance to
this source were $\sim0.4$--1.4\,kpc. The best match to the photometry
using a local extinction correction is a G/K~III, with a G4~III star
the best overall match, which fits with both colours and individual
magnitudes for a Galactic Centre distance.


\begin{table*}
\begin{center}
\begin{tabular}{lccccccc}
\hline
               & $J$ & $H$ & $\ks$ & $\jh$ & $\jk$ & $\hk$ & Candidate Counterparts\\
\hline
all-sky        & $11.00 \pm 0.20$ & $10.73 \pm 0.15$ & $11.24 \pm 0.10$ & $0.70 \pm 0.25$ & $-0.24 \pm 0.22$ & $-0.51 \pm 0.18$ & B0~V \\
Galactic Bulge & $14.40 \pm 0.24$ & $12.98 \pm 0.15$ & $12.87 \pm 0.08$ & $1.42 \pm 0.28$ & $1.53 \pm 0.25$ & $0.11 \pm 0.17$ & K1~III \\
local          & $14.38 \pm 0.27$ & $13.76 \pm 0.12$ & $13.65 \pm 0.04$ & $0.63 \pm 0.29$ & $0.73 \pm 0.27$ & $0.10 \pm 0.13$ & G4~III \\
\hline
\end{tabular}
\caption{The possible spectral types of the bright candidate
counterpart based on analysis of the colours and magnitudes for the
three extinction corrections (assuming a full extinction correction
and a distance of $7.5 \pm 0.45 \,{\rm kpc}$). }
\label{brightposs}
\end{center}
\end{table*}

\subsection{The spectrum}
\label{spec}

\begin{center}
\begin{figure*}
\includegraphics[width=160mm]{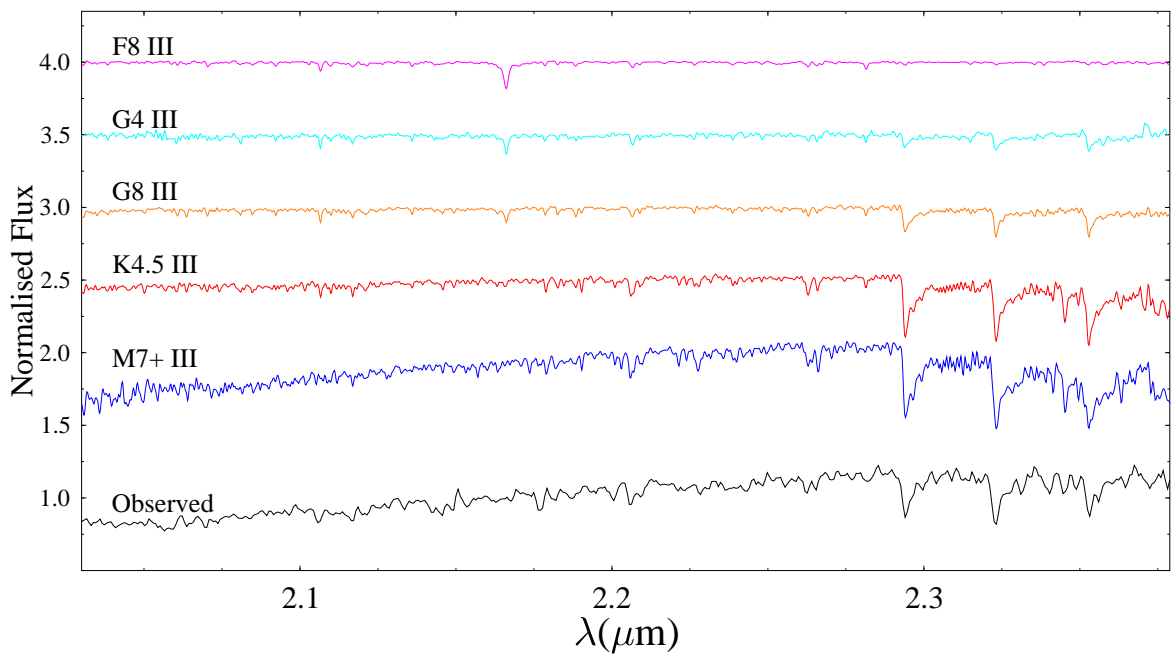}
\caption{Respectively from the bottom to the top: \groj\ spectrum
({\it black}), with the three best solutions to the optimal
subtraction ({\it blue}: M7+~III, {\it red}: K4.5~III, {\it
orange}: G8~III), the best match to the photometry ({\it
cyan}: G4~III) and the latest spectral type excluded by the
optimal subtraction ({\it magenta}: F8~III, top) included for
comparison \citep{wall97}. The standard spectra have been displaced in
amplitude for easier comparison.}
\label{multi}
\end{figure*}
\end{center}

We obtained a spectrum for the brighter of the two stars, shown in
Figures \ref{speccomb} and \ref{multi}.  From the absorption lines in
the spectrum, it appears that this source is a late type giant. There
are strong $^{12}$CO band-heads, some evidence of $^{13}$CO band-heads
and Mg, Ca and Na absorption features.

We performed an optimal subtraction (similar to a $\chi^2$ test where
the difference between the normalised spectra are measured for each
wavelength, the minimum being the best fit) in order to improve our
identification of the spectral type of the counterpart. We used the
spectral library of standards from \citet{wall97} which includes stars
of most spectral types from A--M giants. The resolution of the
standards was R = 3000, so we had to re-sample them to match the lower
resolution of our observed spectrum.

The optimal subtraction enabled us to exclude the possibility that the
counterpart is of spectral type earlier than a G~III type star. This
is based on the absence of CO band-heads from the spectra of these
earlier type stars. The CO band-heads are the most prominent feature
of the observed spectrum and so the subtraction is strongly influenced
by their presence.  Of the G, K and M type spectra, later types of
each (5-8) proved the best fits based on the strength of the CO
features, with G8~III, K4.5~III, and M7+~III proving the best fits for
the respective spectral types as shown in Figure \ref{multi}. The
relatively low resolution of our target spectrum does not allow us to
distinguish between these possibilities in a statistically significant
way.

\brg\ emission was not observed in the spectrum, nor any other
emission lines, indicating the brighter source is not in an accreting
system. Alternatively, if it is, then the surface of the companion
star or the outer regions of the accretion disc if present are not
being heated sufficiently to produce an observable emission line in
our spectrum.

The spectral types indicated by the optimal subtraction are in
agreement with the spectral type as constrained using the photometry
for the Galactic Bulge and local extinction cases. However it is only
the local case that gives a consistent result for all the colours and
magnitudes which also agrees with the spectrum.


\section{Faint star}
\label{faint}

\subsection{Star Properties}

The position of the faint source on the colour-magnitude diagram
(Fig. \ref{colmag} and \ref{colmagjk}) indicates that it is not
actually at the distance of the GC, but is in fact closer. The colour
excess for this star is half that of the GC population in the field,
therefore, we have chosen to apply only half the extinction correction
to this star compared to the brighter counterpart. We have also
applied a distance correction of $3.75 \pm 1 \,{\rm kpc}$ with a
larger error than the derived from the error on the GC distance due to
the uncertainties in the actual distance as judged purely on the
star's position on the colour-magnitude diagram.  Using these values
for the distance and extinction correction, we now examine in detail
the implications for the nature of the fainter star in each of the
three extinction scenarios.  The results are summarised in
Table~\ref{faintposs}.

\subsubsection{Corrected for all-sky extinction}


Applying 0.5 of the all-sky extinction value gives the faint source
intrinsic magnitudes of $J = 14.40 \pm 0.20$, $H = 12.98 \pm 0.15$ and
$\ks= 12.87 \pm 0.10$ which gives it colours of $\jh = 1.42 \pm 0.25$,
$\jk = 1.53 \pm 0.22$ and $\hk = 0.11 \pm 0.18$.

The $\jh$ colour is inconsistent with any spectral type of stars, even
at the limits of the errors. At the extreme of the error bounds, the
$\jk$ is close to that for an M7~III. The $\hk$ colour is consistent
with all spectral types of all stars within the error bounds, but the
best fit is to early K1-4~V/III stars, or G8~I \citep{alle00}. 

The $J$ magnitude is approximately 1.4\,mag fainter than the $H$ and
$\ks$ magnitudes, meaning that the three magnitudes are not consistent
with a single spectral type, but at a distance of $3.75 \pm 1 \,{\rm
kpc}$ \citep{nish06b}, the $H$ and $\ks$ magnitudes are consistent
with a late type B~V star \citep{alle00}.

Considering the spectral types allowed by the colours, the supergiant
would be 6--7 magnitudes brighter than observed if it were at the
distance estimated by the star's position on the colour-magnitude
diagram, or to achieve the observed magnitudes, it would have to be at
a distance $\ge 40\,{\rm kpc}$, on the far side of the
Galaxy. Similarly for the giant, it would be 2--5 magnitudes brighter
if at $3.75 \pm 1 \,{\rm kpc}$, or would have to be at $\ge 7\,{\rm
kpc}$, approximately in the Bulge, in contradiction to its position on
the colour-magnitude diagrams. There is no spectral type that is
consistent with both the colours and magnitudes.


\subsubsection{Corrected for Galactic Bulge extinction}

As above, applying only half of the extinction calculated using the
values from \citet{nish06}, the intrinsic magnitudes of the faint
source become $J = 19.05 \pm 0.26$, $H = 17.58 \pm 0.16$ and $\ks= 17.48
\pm 0.12$, giving the star colours of $\jh = 1.48 \pm 0.31$, $\jk =
1.57 \pm 0.29$ and $\hk = 0.10 \pm 0.21$.

As with the all-sky extinction correction, using the extinction for
the Galactic Bulge, the resulting $\jh$ colour is inconsistent with
all spectral types, and the $\jk$ colour is consistent only with the
late M~III stars at the limit of the errors. The $\hk$ colour allows
for all spectral types within the errors, but the best fits would be
K2/4~V, a K0/1~III or a G3-8~I \citep{alle00}. 

Correcting for half the Galactic centre distance of $3.75 \pm 1 \,{\rm
kpc}$ \citep{nish06b}, as with the previous case, the $J$ magnitude
appears too faint. The $H$ and $\ks$ magnitudes are consistent with a
K5-7~V star, the $J$ magnitude for an M0/1~V star. 

As with the all sky case, the supergiant and giant stars that are
possible matches would be 2--7 magnitudes too bright at the distance
determined from the colour-magnitude diagram. For them to have the
observed magnitudes, they would have to be on the far side of the
Galaxy, or in the Galactic Bulge respectively, but the position of the
star on the colour-magnitude diagrams is inconsistent with this.  A
K4/5~V star is consistent with the $\hk$ colour and $H$ and $\ks$
magnitudes however the limits are most strongly placed by the $H$ and
$\ks$ magnitudes as all spectral types are allowed within the $\hk$
colour errors.

\subsubsection{Corrected for local extinction}
\label{faintloc}

Using the extinction values from Section \ref{extloc}, applying half
of the local extinction, the intrinsic magnitudes of the faint source
are $J = 19.04 \pm 0.29$, $H = 17.99 \pm 0.14$ and $\ks= 17.90 \pm
0.10$, giving the star colours of $\jh = 1.06 \pm 0.32$, $\jk = 1.15
\pm 0.30$ and $\hk = 0.09 \pm 0.17$.

After applying half the local extinction correction, all three colours
of the faint counterpart are consistent with either M~V, K~III or
G--M~I type stars. This is based on the errors in the $\jh$ and $\jk$
colours since, as with the previous cases, the $\hk$ error allows all spectral
types.

All three magnitudes are consistent with a K7/M0~V star when corrected
for a distance of $3.75 \pm 1 \,{\rm kpc}$ \citep{nish06b}.

As with the other two cases, the supergiant and giant spectral types
that match the colours would be 2--7 magnitudes too bright for the
distance suggested by the position of the star on the colour-magnitude
diagrams. A K7/M0~V star is consistent with all the colours and
magnitudes within the errors, and so is the most likely candidate for
the spectral type of this faint candidate counterpart.


\begin{table*}
\begin{center}
\begin{tabular}{lccccccc}
\hline
               & $J$ & $H$ & $\ks$ & $\jh$ & $\jk$ & $\hk$ & Candidate Counterparts\\
\hline
all-sky        & $14.40 \pm 0.20$ & $12.98 \pm 0.15$ & $12.87 \pm 0.10$ & $1.42 \pm 0.25$ & $1.53 \pm 0.22$ & $0.11 \pm 0.18$ & - \\
Galactic Bulge & $19.05 \pm 0.26$ & $17.58 \pm 0.16$ & $17.48 \pm 0.12$ & $1.48 \pm 0.31$ & $1.57 \pm 0.29$ & $0.10 \pm 0.21$ & K4/5~V \\
local          & $19.04 \pm 0.29$ & $17.99 \pm 0.14$ & $17.90 \pm 0.10$ & $1.06 \pm 0.32$ & $1.15 \pm 0.30$ & $0.09 \pm 0.17$ & K7/M0~V \\
\hline
\end{tabular}
\caption{The extinction corrected magnitudes, colours and possible
spectral types of the faint candidate counterpart for the three
extinction corrections (assuming a half extinction correction and a
distance of $3.75 \pm 1 \,{\rm kpc}$).}
\label{faintposs}
\end{center}
\end{table*}


\section{Extinction Discussion}
\label{sec:extinction}

For the all-sky and Galactic Bulge extinction cases, there are clear
problems with the colours and magnitudes obtained after the extinction
correction has been applied. These could be caused by an error in the
extinction correction, be an intrinsic property of the stars, or
result from an error in the original photometry of the sources. The
bright candidate counterpart's colours and magnitudes are consistent
with the overall field population, and with the local surrounding
population suggesting that the resultant problems with the photometry
are not intrinsic to this star. \citet{wang07} observed the bright
candidate counterpart to \groj\ on February $11^{\rm th}$ 2004 using
the PANIC camera on the 6.5-meter Magellan/Baade telescope at Las
Campanas Observatory in Chile. Their reddened magnitudes are similar
to those of this work: $J = 19.21$, $H = 16.16$ and $\ks = 14.69$
$\pm 0.05$, compared to our values of $J = 18.98$, $H = 15.68$ and $\ks
= 14.44$ $\pm 0.02$ (Fig. \ref{colmagjk}).  While there is a slight
offset between the two sets of magnitudes, it is not sufficient to
account for the problems with the photometric classification of the
counterpart. \citet{wang07} use a different photometric calibration to
ours ({\it Z. Wang, private communications}) which can account for the
offset. The fact that they independently observed similar magnitudes
suggests that there is not a calibration error in our data.

This leaves the extinction correction as the most likely cause of the
abnormal un-reddened photometry. As mentioned in Section \ref{ext},
the extinction towards the Galactic Centre is highly variable
\citep[][and many others]{catc90, schu99, dutr03, gosl06}, and there
is evidence to suggest that the behaviour of the extinction law is
also variable not only towards the Galactic Centre but across the
whole sky \citep[][Gosling et al. {\it in prep.}]{nish06,
froe07}. When we applied a localised extinction correction, taking
into account the variation of the extinction law we recovered
consistent colours and magnitudes for the brighter candidate
counterpart that agreed with the spectroscopic identification.

For the fainter candidate counterpart, we only applied an extinction
correction of half that used for a star at the GC based on the
position of the star on the colour-magnitude diagram. 
As with the bright candidate counterpart, using either an all-sky
extinction correction or a Galactic Bulge correction were not able
to recover a consistent spectral type (see Section
\ref{faint}). This was again caused by the $J$ magnitude
seeming to be anomalously faint in comparison to the $H$ and
$\ks$. The localised extinction correction that takes into account a
variation in the wavelength dependence of the extinction recovered
colours and magnitudes that were all consistent with a K7/M0~V star at
a distance of $3.75 \pm 1 \,{\rm kpc}$.

For both the bright and faint stars, we find that the $J$ magnitude
measured is anomalously faint for both an all-sky and general Galactic
Bulge extinction correction. For these cases, this prevented
identification of a specific spectral type of the counterpart.  We
calculated a local extinction correction, which takes into account
variations in the extinction law across the Galactic Bulge, which gave
an extinction law of $\alpha = 3.23 \pm 0.01$. This is significantly
steeper than the extinction law was previously thought to be in this
region \citep{riek85,card89,schu99,dutr03,inde05,nish06} but would
explain why in the all-sky and Galactic Bulge extinction cases we find
that it is the $J$ magnitude that is anomalously faint, as such a high
value for the extinction law means that the extinction will be
proportionally underestimated at shorter wavelengths in such cases.

The local extinction case is the only one to produce colours and
magnitudes that are all consistent with a single (or very small range)
of spectral types, and in the case of the bright candidate counterpart, this
spectral classification is in general agreement with the spectral
types allowed based on the comparison of the observed spectrum to
standard stars. We feel that although it is a departure from the accepted
results for extinction corrections, this does result in accurate
photometry for the target stars, and as we will show in Gosling et
al. {\it in prep}, and has been shown in \citet{nish06b, fitz07} and
\citet{froe07} is a phenomenon not limited to this one area, but is
present across the entire Galactic plane.


\section{Counterpart Discussion}

\subsection{Bright Counterpart}
\label{discbright}

We obtained a spectrum of the brighter candidate, and the prominent CO
and metal absorption features indicate that it is a late type giant
star. The optimal subtraction that we performed narrowed it down to a
late G, K or M giant, however the signal-to-noise of the spectrum was
insufficient to be able to improve on this. Combining this with the
photometry (using the local extinction correction) the bright
candidate counterpart is most likely a late G/K~III star.

 Using this as a constraint, we can estimate some of the physical
properties of the system.  We use a NS mass of $M_{\rm x} = 1.4 \pm
0.1\, {\rm M_{\odot}}$, a companion G/K~III mass of $M_{\rm c} = 2.5
\pm 0.5\, {\rm M_{\odot}}$ \citep{alle00} and the measured orbital
period of $P_{{\rm orb}} = 11.8337 \pm 0.0013\, {\rm days}$
\citep{fing96}.  From this we calculate the system would have a
semi-major axis of $a = 31 \pm 4\, {\rm R_{\odot}}$, and the Roche
lobe of the donor star would be $R_{\rm RL} = 12 \pm 4\, {\rm
R_{\odot}}$.  In this case, if the bright star is the counterpart,
then the $6 - 15\,{\rm \Rsol}$ radius of a G/K~III giant
\citep{alle00} is on the limits for accretion to be by Roche lobe
overflow.  A star of this mass is also inconsistent with the
mass-function measured by \citet{fing96}, unless the inclination angle
$i < 4^{\circ}$, i.e.\ we are observing the system nearly pole-on. 

No \brg\ line was observed in the spectrum of this candidate
counterpart, nor any other emission lines. Therefore if this source is
the counterpart to \groj\ then the surface of the star/outer accretion
disc is not being illuminated and heated by the X-rays from the
compact object, or if so the heating is not strong enough to cause
detectable line emission. The source may not be in an accreting phase;
if so the observed X-rays could be the residual cooling of the NS
surface, assuming the quiescent flux measured by \citet{wijn02,
wang07}. If so, and this star is the companion, this would be in
agreement with the theory put forward by \citet{cui97} that \groj\ is
in the propeller regime which is inhibiting accretion when it is in
its quiescent state.

\subsection{Faint Counterpart}
\label{discfaint}

Using the local extinction correction, and a distance of $3.75 \pm 1
\,{\rm kpc}$ the most likely spectral type for the star is
K7/M0~V. This was the only result that was in agreement for all
colours and magnitudes.

A K7/M0~V star has a mass of $M_{\rm c} = 0.5 \pm 0.1\, {\rm
M_{\odot}}$ and radius of $R_{\rm c} = 0.7 \pm 0.1\, {\rm R_{\odot}}$
\citep{alle00}. Performing the same calculation as in Section
\ref{discbright}, using the parameters from \citet{fing96} gives an
orbital separation of $a = 28 \pm 4\, {\rm R_{\odot}}$, and the Roche
lobe of the donor star would be $R_{\rm RL} = 9 \pm 4\, {\rm
R_{\odot}}$. Mass transfer in such a system would be by wind
accretion. Such a system would have an inclination angle, $i <
9^{\circ}$, still very unlikely.

The magnetic field strength estimated by \citet{cui97} ($10^{11} {\rm
G}$), requires a mass accretion rate $< 2.32 \times 10^{15}\, {\rm
kg\, s^{-1}}$ for the Alfv\'{e}n radius to be greater than the
co-rotation radius for this system, as required for the propeller
effect. This is about 4 orders of magnitude greater than typical
wind-accretion rates, so if the system is wind-fed then the observed
quiescence could definitely be caused by the propeller effect as
theorised by \citet{cui97}. However it is very unlikely that X-ray
emission would be observed from a wind-accreting system containing a
late main-sequence star as the stellar wind would result in very
little mass transfer.

The distance used in calculations for this star, $3.75 \pm 1 \,{\rm
kpc}$, is slightly lower than the $L_{\rm Edd}$ distance as calculated
by \citet{gile96}. It is also at odds with the measured column density
to the source that indicates that it should be closer to the GC
\citep[][and others]{dota96a, gile96, muno07}. However, the extinction
correction that has provided the best corrected matches for the
photometry (and compared to the spectroscopy in the bright candidates
case) has a very steep extinction law, $\alpha = 3.23 \pm 0.01$. If
this steep law extends to X-ray wavelengths, it would cause an
over-estimation of the source distance based on the column density
measurement.

This star was too faint to obtain a spectrum with the integration
times used for our 2003 VLT/ISAAC spectroscopy.  In order to
definitively rule out this star as the counterpart to \groj, high S/N
K-band spectroscopy, with significantly higher integration times, will
be required.

\subsection{Mass Function Constraints}
\label{mass}

\begin{center}
\begin{figure}
\includegraphics[width=\columnwidth]{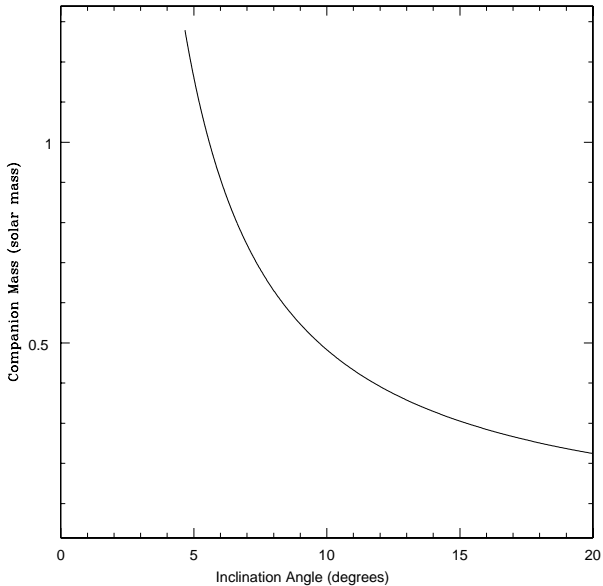}
\caption{Companion mass as a function of inclination angle based on
the orbital constraints from \citet{fing96}. Calculated for $M_x = 1.4
\Msol$.}
\label{angle}
\end{figure}
\end{center}

The mass-function for \groj\ determined by \citet{fing96} is
$f_{\rm x}(M) = 1.36 \times 10^{-4} {\rm M_{\odot}}$. Using this
in conjunction with the $P_{\rm orb}$ \citep[also from][]{fing96}
we calculated the expected mass of the companion as a function of
the inclination angle, assuming a neutron star mass of $M_{\rm x}
= 1.4 {\rm M_{\odot}}$ (Fig. \ref{angle}). From this we find
that unless we are observing the system almost pole-on
(inclination angle $< 15^{\circ}$), the companion will have
$M_{\rm C} \ltsim 0.3\,\Msol$, in agreement with the calculations
of \citet{bild97}.


The mass function is still the strongest constraint on the
properties of the counterpart. Neither of the stars detected in
the X-ray error circle are consistent with it unless the system is
almost pole-on, in which case almost any spectral type is possible
(Fig. \ref{angle}). \citet{fing96} measured the orbital period
by fitting pulse phases, caused by Doppler shifting due to orbital
motion, which would be un-observable if the system were pole-on.
Both the candidates described above would require the system to be
almost pole-on and so are unlikely to be the true counterparts.

Assuming the system is not pole-on, the mass-function places an
approximate upper limit on the mass of the counterpart at
$\sim\!0.3\,\Msol$ ($i > 15^{\circ}$), which equates to an M3 or later
main sequence star \citep{alle00}, or an evolved star that has lost a
significant portion of its mass \citep{rapp97}. Both of these
scenarios rule out Roche Lobe overflow as the method of production of
the currently observed X-ray emission in \groj\ as neither would fill
their Roche Lobes. Thus, the observed X-ray emission is either
residual cooling of the surface, or wind accretion and as such in
quiescence \groj\ would have a minimal disc unlikely to affect the
observed magnitude of the counterpart.

If the counterpart were a low-mass main sequence star, at the distance
of the GC, it will be too faint to be detected in our ISAAC images
($K_{{\rm S app}} \gtsim 20$). It would require deep, very-high
spatial resolution observations to reveal such a star, however such a
low-mass star would never be able to produce a sufficiently strong
wind to accrete enough material to explain the previous periods of
outburst observed in the system. This will also be true of the faint
candidate counterpart and as such it is unlikely that this star can be
the counterpart.

\citet{rapp97} calculated that if the counterpart to \groj\ was an
evolved star that had had the majority of its enveloped stripped
through binary interaction it would have a $\ks$-band magnitude most
likely of 14.1 with a range of 13.5-14.5 (95\% confidence limit). The
magnitude limit of our survey is $\ks=20$ so even with the highest
extinction of the all-sky value, we would have detected such a
star. The bright star in our survey, when corrected for the best,
local extinction correction, has a magnitude of $\ks= 13.65 \pm 0.04$
within the confidence limits of this value. \citet{rapp97} give a
radius of $R_{\rm c} = 6.2\,\Rsol$ ($6 - 9\,\Rsol$ 95\% confidence
range) for such a star as it has lost part of its envelope. This star
would be even less likely to be accreting by Roche lobe overflow,
although it is still possible within the errors. We do not observe
\brg\ emission in the spectrum of this star, but as it is in
quiescence, it is very likely that the emission is too faint to
detect. However, this star would have an extremely low inclination $i
< 4^{\circ}$, even lower than the constraints from \citet{rapp97} $i =
18^{\circ}$ ($7^{\circ}\!-\!22^{\circ}$, 95\% confidence limit) but as
this counterpart has lost a considerable amount of its envelope, it
could have a lower mass than stated without having significantly
changed its observed spectral type and therefore a slightly larger
inclination angle. This low inclination angle is unlikely in a single
system, but as \citet{rapp97} point out, by 1997, about 100 X-ray
binaries had been studied in detail so it is reasonable to expect that
there should be a few systems with very low inclinations.


\section{Conclusions}

We detected two sources within the X-ray error circle of \groj. The
bright source has a position $R.A. = 17^{\rm h}44^{\rm m}33\fs07$,
$Dec. = -28^{\circ}44\arcmin26\farcs89$ and the faint source has
position $R.A. = 17^{\rm h}44^{\rm m}33\fs16$, $Dec. =
-28^{\circ}44\arcmin27\farcs41$ with $0\farcs1$ errors.

We calculated the extinction towards these sources using three
methods. A canonical all-sky value using the extinction law from
\citet{riek85}, a Galactic Bulge value using the extinction law
measured in \citet{nish06} and a local value calculated using the
colour-excesses of the stars within 20\arcsec\ of the sources (see
Section \ref{ext}). When correcting for these three extinction cases,
only the local value produced colours and magnitudes that were all
consistent with each other and a single, or small range, of spectral
types (see Sections \ref{brightloc} and \ref{faintloc}), and
consistent with the spectroscopic identification for the bright
counterpart (see Section \ref{spec}). This local extinction value has
an extinction law relation of $\alpha = 3.23 \pm 0.01$ ($A_{\lambda}
\propto \lambda^{-\alpha}$), much steeper than the previously accepted
values. This explains why in the all-sky and Galactic Bulge extinction
cases the $J$ magnitude is always faint compared to the $H$ and $\ks$
magnitudes as the $J$ extinction will be proportionally underestimated
compared to the extinction at these two wavelengths.

The mass-function for \groj, for an inclination of $i \geq 15\degr$,
constrains the mass of the counterpart to be $M < 0.3\,\Msol$. For
this mass, the counterpart would have to be either an M3+~V
\citep{alle00} or an evolved star that has been stripped of mass
\citep{rapp97}. An M3+~V is too faint to be detected in this survey,
assuming a Galactic Centre distance of $7.5 \pm
0.45\,{\rm kpc}$ and that the extinction is consistent with the local
extinction case, and to be detected would require an observation with approximate limits
of $J = 26.5$, $H = 23$ and $\ks = 22$, 4 magnitudes deeper than the
observations of this work. \citet{rapp97} give limits for the
magnitude of the likely counterpart to \groj\ if it is a stripped
giant, and such a star would have been detected in the observations
for this work. In either case, the counterpart would not fill its
Roche Lobe, and mass transfer in the system, if there is any, would be
by wind accretion. In this case, the magnetic field strength measured
by \citet{cui97} would mean that the propeller effect was inhibiting
accretion explaining the quiescent state of \groj\ as observed since
1997.
%
%
%
%

Photometry of the faint star indicates that its most likely spectral
type is K7/M0~V. Based on comparison with the mass-function, it would
require the system to be almost pole-on ($i < 9^{\circ}$).  The
distance of this source \citep[$3.75 \pm 1 {\rm kpc}$, see
Fig. \ref{colmag} \& ][]{nish06b} is smaller than suggested by both
the $L_{\rm Edd}$ distance calculated by \citet{gile96}, and the
measured column densities which suggest the source is close to the GC.
However, if the steep extinction law slope of the local extinction
extends to X-ray wavelengths it could cause an over-estimation of the
distance to the X-ray source using only the column density. This faint
source thus cannot be ruled out as the counterpart, but further data
are required, especially IR spectroscopy, to investigate this
possibility.

Based on photometry and spectroscopy, the bright candidate is most
likely a late type G/K~III star. The mass-function for the \groj\
system \citep{fing96} means that if this is the counterpart, the
system will be almost pole-on to the observer ($i < 4^{\circ}$). The
spectrum shows no sign of \brg, so, if this source is indeed the
counterpart, the illumination of the outer accretion disc or surface
of the star is insufficient to produce an observable emission
line. The constraints on the observed magnitude of the counterpart
that are placed by the calculations of \citet{rapp97} agree with the
un-reddened magnitude of the bright counterpart (within the 95\%
confidence limits) if it is an evolved giant that has lost a
significant fraction of its envelope. If this is the case, and this
star is the true counterpart, the
mass-loss would mean that a higher inclination angle is also
possible. As stated in \citet{rapp97}, by 1997 over 100 X-ray binaries
had been studied in detail so it is plausible that a system with such a
low inclination angle could by now be included in the sample.

Although the evidence is in favour of the brighter of the two
candidate counterparts being the companion to the \groj\ X-ray source,
without detection of a \brg\ line in the spectrum, or further
observations to rule out a fainter source, a definitive answer cannot
be given as to the nature of the companion.

\section{Acknowledgements}

AJG would like to thank the UK Particle Physics and Astronomy Research
Council for his studentship. SAF acknowledges travel support provided
by UNSW@ADFA and the Astronomical Society of Australia. This paper is
based on observations made with the ESO VLT at Paranal under imaging
programme ID 071.D-0377(A) and spectroscopic programme ID
075.D-0361(A).  We would also like to thank the anonymous referee
for the helpful comments made.



\label{lastpage}

\end{document}